\documentclass{an}
\usepackage{graphicx}
\usepackage{html}
\usepackage{url}
\usepackage{times}
\newcommand{\EQ}{\begin{equation}}
\newcommand{\EN}{\end{equation}}
\newcommand{\EQA}{\begin{eqnarray}}
\newcommand{\ENA}{\end{eqnarray}}
\newcommand{\eq}[1]{(\ref{#1})}
\newcommand{\EEq}[1]{Equation~(\ref{#1})}
\newcommand{\Eq}[1]{Eq.~(\ref{#1})}
\newcommand{\Eqs}[2]{Eqs~(\ref{#1}) and~(\ref{#2})}

\newcommand{\Fig}[1]{Fig.~\ref{#1}}
\newcommand{\FFig}[1]{Figure~\ref{#1}}

\newcommand{\bra}[1]{\langle #1\rangle}

\newcommand{\meanemf}{\overline{\mbox{\boldmath ${\cal E}$}} {}}
\newcommand{\meanAA}{\overline{\bf{A}}}
\newcommand{\meanBB}{\overline{\bf{B}}}
\newcommand{\meanJJ}{\overline{\bf{J}}}
\newcommand{\meanUU}{\overline{\bf{U}}}
{}
{}
{}
{}
{}
{}
{}
\newcommand{\nnn}{\hat{\mbox{\boldmath $n$}} {}}

\newcommand{\zzz}{\mbox{\boldmath $z$} {}}

\newcommand{\uu}{{\bf{u}}}
\newcommand{\BB}{{\bf{B}}}
\newcommand{\JJ}{{\bf{J}}}
\newcommand{\jj}{{\bf{j}}}
\newcommand{\AAA}{{\bf{A}}}
\newcommand{\aaaa}{{\bf{a}}}
\newcommand{\bb}{{\bf{b}}}

\newcommand{\SSS}{{\bf{S}}}

\newcommand{\nab}{\mbox{\boldmath $\nabla$} {}}

\newcommand{\oo}{\mbox{\boldmath $\omega$} {}}

\newcommand{\sgn}{{\rm sgn}  \, {}}

\newcommand{\dd}{{\rm d} {}}

\newcommand{\ea}{{\rm et al.\ }}

\def\onethird{{\textstyle{1\over3}}}

\newcommand{\kG}{\,{\rm kG}}

\newcommand{\Mx}{\,{\rm Mx}}

\newcommand{\astroph}[1]{\htmladdnormallink{astro-ph/#1}{http://xxx.lanl.gov/abs/astro-ph/#1}}

\newcommand{\yjgr}[3]{ #1, {JGR,} {#2}, #3}
\newcommand{\ysol}[3]{ #1, {Sol. Phys.,} {#2}, #3}
\newcommand{\yapj}[3]{ #1, {ApJ,} {#2}, #3}
\newcommand{\yapjl}[3]{ #1, {ApJ,} {#2}, #3}

\newcommand{\yan}[3]{ #1, {AN,} {#2}, #3}

\newcommand{\yana}[3]{ #1, {A\&A,} {#2}, #3}

\newcommand{\yjfm}[3]{ #1, {JFM,} {#2}, #3}
\newcommand{\ypf}[3]{ #1, {Phys. Fluids,} {#2}, #3}
\newcommand{\ypp}[3]{ #1, {Phys. Plasmas,} {#2}, #3}

\newcommand{\yprl}[3]{ #1, {PRL,} {#2}, #3}

\newcommand{\ysph}[3]{ #1, {Solar Phys.,} {#2}, #3}
\newcommand{\ypr}[3]{ #1, {Phys. Rev.,} {#2}, #3}

\newcommand{\yjour}[4]{ #1, {#2}, {#3}, #4}

\newcommand{\yproc}[5]{ #1, in {#3}, ed. #4 (#5), #2}

\newcommand{\papj}[1]{ #1, {ApJ,} (in press)}

\newcommand{\pjour}[2]{ #1, {#2,} (in press)}

\begin{document}
\lhead[\thepage]{A.~Brandenburg \& W.~Dobler: Solar and stellar dynamos}
\rhead[Astron. Nachr./AN~{\bf XXX} (200X) X]{\thepage}
\headnote{Astron. Nachr./AN {\bf 32X} (200X) X, XXX--XXX}

\title{Solar and stellar dynamos -- latest developments}
\author{Axel Brandenburg\inst{1}
  \and Wolfgang Dobler\inst{2}
}

\institute{
NORDITA, Blegdamsvej 17, DK-2100 Copenhagen \O, Denmark \and
Kiepenheuer Institute for Solar Physics, Sch\"oneckstr. 6, 79104 Freiburg, Germany
}

\date{\today,~ $ $Revision: 1.16 $ $}

\abstract{
Recent progress in
the theory of solar and stellar dynamos is reviewed. Particular emphasis
is placed on the mean-field theory which tries to describe the collective
behavior of the magnetic field. In order to understand solar and stellar
activity, a quantitatively reliable theory is necessary. Much of the
new developments center around magnetic helicity conservation which is
seen to be important in numerical simulations. Only a dynamical, explicitly time dependent
theory of $\alpha$-quenching is able to describe this behavior correctly.
}

\correspondence{brandenb@nordita.dk}

\maketitle

\section{Introduction}

Starspot activity is presumably driven by some kind of
dynamo process. Many stars show magnetic field patterns extending
over scales of up to $30^\circ$ in diameter. The commonly used tool
to model such magnetic activity is the mean-field dynamo.
Although mean-field theory has been used over several decades
there have recently been substantial developments concerning the
basic nonlinearity of dynamo theory. It is the purpose of this
review to highlight these recent developments in the light
of applications to stars.

\section{Stellar dynamos: spots and cycles}

We usually think of star spots as rather extended dark
and strongly magnetized areas on a stellar surface.
Observable spots are
much bigger than sunspots. They may in fact be so big that
the spots themselves have sometimes been identified with solutions of
the mean-field dynamo equations (e.g., Moss \ea 1995).
This is in contrast to the much smaller sunspots which instead
show collective
behavior in that sunspot pairs have a systematic orientation
and preferential location which changes with the solar cycle.

The working hypothesis is that extended star spots are just the
extremes of a broad range of possibilities from small to large
spots. Stellar parameters can change over a considerable range and there is scope
that different types of behavior can be identified with different
solutions of the mean-field dynamo equations. Very exiting is the
possibility of nonaxisymmetric solutions, possibly with cyclic
nonmigratory alternations of their polarity (the so-called
flip-flop effect, see Jetsu \ea 1999 and references therein).

Already among the more solar-like stars there is a lot to be
learned about the dependence of the period of the activity cycle
on rotation rate and spectral type (cf.\ Baliunas et al.\ 1995). An interesting possibility
is the suggestion that stellar activity behavior may change with
age (Brandenburg, Saar \& Turpin 1998). The very young and
more active stars show extremely long cycles (3--4 orders of
magnitude longer than the rotation period) whilst older inactive stars like
the sun show shorter cycle periods that are just a few hundred
times longer than the rotation period. These different types of
behavior can be classified according to their location in the
Rossby number versus frequency ratio diagram
(Saar \& Brandenburg 1999).

In order to make progress in understanding these various
possible behaviors it is crucial to work with a reliable theory
that has predictive power.
Mean-field dynamo theory has frequently been used as a rather
arbitrary theory.
Being based on some ad hoc assumptions, much of its
predictive power is questionable.
Particular controversy was caused by the ill-known contributions of
small scale fields which may catastrophically quench the $\alpha$-effect
(Vainshtein \& Cattaneo 1992, Kulsrud \& Andersen 1992), which is
thought to be responsible for driving the large scale field.
However, significant advances in recent years are now beginning to
shed some light on apparently conflicting earlier results on what
the final saturation field strength will be.
It is likely that progress will come about in two stages.
In the first stage we will have to make sure that mean-field theory works
correctly in the parameter regime that can be tested using simulations. 
In the second stage we have to extrapolate the theory from the
regime that is tested numerically to the regime that is of
astrophysical interest. At the moment we are still struggling
with the first objective.

\section{Mean-field theory: it's all about quenching}

As far as the selection of different modes of symmetry is concerned,
there has been some partial success in finding agreement between
simulations and linear dynamo theory. We mention here the results of
local simulations of accretion discs in a shearing box approximation:
changing the upper and lower boundary conditions from a normal field
(``pseudo-vacuum'')
condition to a perfect conductor condition changes the
behavior from an oscillatory mode with symmetric field about
the equator to a non-oscillatory mode with antisymmetric field
about the equator. The same change of behavior is also seen in mean-field
models using the same cartesian geometry.  This result, which has been
described in more detail in earlier papers (e.g.\ Brandenburg 1998), lends
some support to the basic idea of using mean-field theory to describe the
results of simulations of hydromagnetic turbulence under the influence
of rotation and stratification and in the presence of boundary conditions.

More serious concern comes from the effects of nonlinearity. Broadly
speaking, nonlinearity has to do with strong magnetic fields,
where the magnetic energy density approaches the kinetic energy density
of the turbulence. 
At the bottom of the solar convection zone, the corresponding
equipartition field strength is $B_{\rm eq}=4\ldots8\kG$.
On the one hand, magnetic fields of this strength may actually be
required for the dynamo to operate. Babcock (1961) and Leighton (1969)
proposed that magnetic fields of this strength will become buoyant
and produce, under the influence of the Coriolis force, a systematic
tilt as flux tubes emerge at the surface to form a sunspot pair.
In many ways magnetic buoyancy is similar to thermal buoyancy and
both lead to an $\alpha$-effect (Parker 1955,
Steenbeck, Krause \& R\"adler 1966).
However, Piddington (1972) has argued that, when the
magnetic field approaches the equipartition value, it would be impossible
to entangle and diffuse the magnetic field. This led later to the idea
of catastrophic suppression of turbulent magnetic diffusivity
(Cattaneo \& Vainshtein 1991) and, by analogy, to the proposal of
catastrophic suppression of the $\alpha$-effect
(Vainshtein \& Cattaneo 1992). Simulations of
Tao \ea (1993) and Cattaneo \& Hughes (1996) show that
in the presence of an imposed magnetic field, $\BB_0$, the
$\alpha$-effect depends on $\BB_0$ like
\EQ
\alpha={\alpha_0\over1+R_{\rm m}\BB_0^2/B_{\rm eq}^2},
\label{catastrophic}
\EN
where $R_{\rm m}$ is the magnetic Reynolds number which is large
($10^{8\ldots9}$ for the sun). Thus, for equipartition field
strengths, $B_0\sim B_{\rm eq}$, $\alpha$ would be 8 to 9
orders of magnitude below the kinematic (unquenched) value
$\alpha_0$, i.e.\
\EQ
\alpha\sim\alpha_0 R_{\rm m}^{-1}\rightarrow0
\quad\mbox{as}\quad R_{\rm m}\rightarrow\infty.
\EN

Over the past ten years there has been an increasing amount of activity
in trying to understand the value of $\alpha$ in the nonlinear regime.
Work by Gruzinov \& Diamond (1994, 1995, 1996) and Bhattacharjee \& Yuan
(1995) has basically confirmed \Eq{catastrophic}. Gruzinov \& Diamond (1994)
did find however that the turbulent magnetic diffusivity is only quenched
in two-dimensional configurations,
which was exactly the case considered numerically by
Cattaneo \& Vainshtein (1991). Although Gruzinov \& Diamond (1994)
did agree with the conclusion of catastrophic $\alpha$-quenching,
they found actually a slightly different form of \Eq{catastrophic}, which
can be written as
\EQ
\alpha={\alpha_0+R_{\rm m}\,\eta_{\rm t}\mu_0\meanJJ\cdot\meanBB/B_{\rm eq}^2
\over1+R_{\rm m}\meanBB^2/B_{\rm eq}^2},
\label{catastrophicJ}
\EN
where $\meanJJ=\nab\times\meanBB/\mu_0$ is the mean current density and
$\mu_0$ the vacuum permeability. Obviously,
when the mean field is spatially uniform, $\meanBB=\BB_0=\mbox{const}$,
then $\meanJJ=0$ and \Eqs{catastrophic}{catastrophicJ} agree with each other.

\begin{figure}[t!]\centering\includegraphics[width=0.5\textwidth]{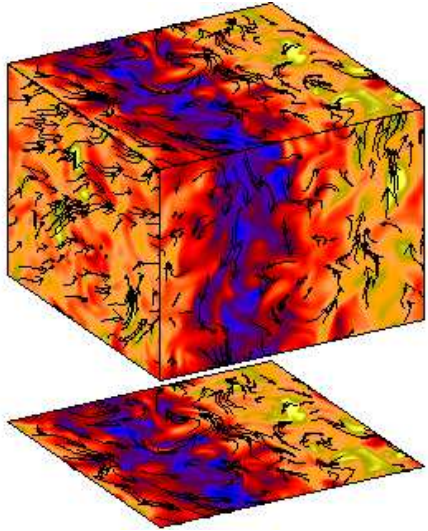}\caption{
Visualization of the magnetic field in a three-dimensional simulation
of helically forced turbulence. The turbulent magnetic field is
modulated by a slowly varying component that is force-free.
}\label{FBeltrami}\end{figure}

In real astrophysical bodies $\alpha$ will always be a tensor
(e.g., Steenbeck \ea 1966, R\"udiger \& Kitchatinov 1993,
Rogachevskii \& Kleeorin 2001).
However, much of the work on the nonlinear $\alpha$-effect comes from
considering periodic boxes where the $\alpha$ tensor can indeed be isotropic
(e.g.\ Field, Blackman \& Chou 1999). There is a priori no reason to
assume that the $\alpha$-effect in a periodic box is different
from that in a nonperiodic box. Furthermore, periodic boxes are
conceptually and computationally significantly easier than
boxes with boundaries. If no large scale field is imposed,
helical turbulence can still drive a large scale field which
itself is helical. A prototype of such a field is
\EQ
\meanBB=B_0\left(\sin k_1z,\;\cos k_1z,\;0\right),
\label{Beltrami}
\EN
where $k_1=2\pi/L$ is the smallest wavenumber in a box of size $L^3$.
Other directions and additional phase shifts are possible;
see Brandenburg (2001, hereafter B01); see \Fig{FBeltrami}.
Animations of $x$, $y$, and $z$ slices of the generated magnetic field,
together with the corresponding power spectra of kinetic and magnetic
energy, as well as magnetic helicity (normalized by $k/2$) are shown in
the attached Movies 1--3.

A helical field of the form \eq{Beltrami} is called a Beltrami field.
The current density of such a field no longer vanishes; in fact,
$\mu_0\meanJJ=k_1\meanBB$, so $\meanJJ$ and $\meanBB$ are aligned and
\EQ
\mu_0\meanJJ\cdot\meanBB=k_1\meanBB^2=\mbox{const}.
\label{JBm}
\EN
Thus, in the large magnetic Reynolds number limit, \Eq{catastrophicJ}
becomes
\EQ
\alpha={\cal O}(\eta_{\rm t}k_1)\neq0
\quad\mbox{as}\quad R_{\rm m}\rightarrow\infty.
\EN
This highlights the great ambiguity in concluding anything about
$\alpha$-quenching from oversimplified experiments.
[In the discussion above we have assumed that $\alpha_0>0$.
If $\alpha_0<0$, as is the case in B01, then {\it both} $\alpha$ and
$\meanJJ\cdot\meanBB$ are also negative and \Eq{catastrophicJ}
is unchanged.]

\section{Relation to magnetic helicity}

There is a strong connection between $\alpha$-quenching and
magnetic helicity conservation. Again, we consider the case of
a periodic box, for which it is easy to show that the magnetic helicity
\EQ
  H \equiv \int \AAA\cdot\BB\,\dd V = \bra{\AAA\cdot\BB}V
\EN
is perfectly
conserved in the limit of infinite magnetic Reynolds number.
Here, $\AAA$ is the magnetic vector potential, so the
magnetic field is $\BB=\nab\times\AAA$, and angular brackets denote
volume averages over the full box.
If we start with a very weak seed magnetic field, the initial
magnetic helicity must also be small and will therefore always remain
small if $H$ is conserved.
Thus, a large-scale helical field of the form \eq{Beltrami} is only
compatible with conservation of magnetic helicity if there is an equal
amount of magnetic helicity of the opposite sign in the small scales, i.e.\
\EQ
\bra{\AAA\cdot\BB}\equiv
\bra{\meanAA\cdot\meanBB}+\bra{\aaaa\cdot\bb}=0
\quad\mbox{(early times)},
\label{helcons}
\EN
where $\AAA$ and $\BB$ have been
split up into their mean and fluctuating components,
\EQ
\AAA=\meanAA+\aaaa,\quad\BB=\meanBB+\bb.
\EN
Overbars refer to the mean-field obtained by horizontal or
azimuthal averaging, for example.
\EEq{helcons} is a crucial condition that must be obeyed by any mean-field
theory in the large $R_{\rm m}$ limit on time scales {\it shorter} than the
resistive time scale.

To our knowledge, there have been two approaches to incorporate magnetic
helicity conservation into mean-field dynamo theory. One is to express
the mean turbulent electromotive force, $\meanemf\equiv\overline{\uu\times\bb}$,
as a divergence term (Bhattacharjee \& Hameiri 1986, see also Boozer 1993)
and the other is to
modify the feedback onto the $\alpha$-effect such that \Eq{helcons} is
satisfied on short enough time scales. The latter approach goes back to
Kleeorin \& Ruzmaikin (1982) and Kleeorin \ea (1995), and has recently
been revived by Field \& Blackman (2002), Blackman \& Brandenburg (2002),
and Subramanian (2002).
In the following we briefly outline the basic idea.

All we know is that in a closed or periodic domain the magnetic helicity
evolves according to
\EQ
{\dd\over\dd t}\bra{\AAA\cdot\BB}=-2\eta\mu_0\bra{\JJ\cdot\BB},
\label{helicity_eqn}
\EN
where $\bra{\AAA\cdot\BB}V$ and $\bra{\JJ\cdot\BB}V$ are
magnetic and current helicities, respectively, and $V$ is the volume.
At the same time we have to have some theory for the evolution of the mean magnetic
field. The mean-field $\alpha\Omega$ dynamo equations can be written in the form
\EQ
{\partial\meanBB\over\partial t}=\nab\times
\left[\meanUU\times\meanBB+\alpha\meanBB
-(\eta+\eta_{\rm t})\mu_0\meanJJ\right].
\label{fullset1}
\EN
 From this we can construct the evolution equation for the magnetic helicity
of the mean field,
\EQ
{\dd\over\dd t}\bra{\meanAA\cdot\meanBB}=
2\bra{\meanemf\cdot\meanBB}-2\eta\mu_0\bra{\meanJJ\cdot\meanBB},
\label{dABdt}
\EN
where
\EQ
\meanemf=\alpha\meanBB-\eta_{\rm t}\mu_0\meanJJ,
\label{emf}
\EN
is the mean turbulent electromotive force under the assumption of isotropy.
Subtracting \Eq{dABdt} from \Eq{helicity_eqn}, we obtain the evolution equation
for the magnetic helicity of the fluctuating field,
\EQ
{\dd\over\dd t}\bra{\aaaa\cdot\bb}=
-2\bra{\meanemf\cdot\meanBB}-2\eta\mu_0\bra{\jj\cdot\bb}.
\label{dabdt}
\EN
This equation has to be solved simultaneously with the usual mean field equation.
At the moment, however, it is not yet fully coupled to the mean field equation.
In fact, any kind of coupling, for example of the form
\EQ
\alpha=\alpha\left(\bra{\aaaa\cdot\bb},\bra{\jj\cdot\bb}\right)
\EN
would suffice. A similar relation could in principle also be applied to
the turbulent magnetic diffusivity, $\eta_{\rm t}$. However, in contrast to $\eta_{\rm t}$,
$\bra{\aaaa\cdot\bb}$ and $\bra{\jj\cdot\bb}$ are pseudo-scalars
and change sign when $z$ is changed to $-z$. Therefore, only quadratic constructions
of the form $\bra{\aaaa\cdot\bb}^2$ and $\bra{\jj\cdot\bb}^2$ could,
at least in principle, enter into the feedback of $\eta_{\rm t}$.

In an isotropic periodic box we have
\EQ
\bra{\jj\cdot\bb}=k_{\rm f}^2\bra{\aaaa\cdot\bb},
\EN
where $k_{\rm f}$ can be defined by this relation as the typical
wavenumber of the fluctuating field. Secondly, we use the relation
(Pouquet, Frisch \& L\'eorat 1976)
\EQ
\alpha=-\onethird\tau\bra{\oo\cdot\uu}+\onethird\tau\bra{\jj\cdot\bb}/\rho_0
\label{residual_alpha}
\EN
for the residual $\alpha$-effect. This relation describes a fundamental
form of $\alpha$-quenching, but there could still be additional feedback
onto $\bra{\oo\cdot\uu}$ and $\bra{\uu^2}$
(Rogachevskii \& Kleeorin 2001, Kleeorin \ea 2002), which is ignored here.
With these relations, the equation for
$\alpha$ becomes explicitly time-dependent,
\EQ
{\dd\alpha\over\dd t}=-2\eta_{\rm t0}k_{\rm f}^2\left(
{\alpha\bra{\meanBB^2}-\eta_{\rm t}\mu_0\bra{\meanJJ\cdot\meanBB}
\over B_{\rm eq}^2}+{\alpha-\alpha_0\over R_{\rm m}}\right),
\label{fullset2}
\EN
where $\alpha_0=-\onethird\tau\bra{\oo\cdot\uu}$ is the kinematic value of $\alpha$.
Here we have expressed the correlation time $\tau$ in terms of $B_{\rm eq}$ using
$\eta_{\rm t}=\onethird\tau\bra{\uu^2}$ and $B_{\rm eq}^2=\mu_0\rho_0\bra{\uu^2}$.
The full set of equations to be solved comprises thus \Eqs{fullset1}{fullset2}.

A detailed analysis of this set of equations was given by Field \& Blackman (2002)
for the case of the $\alpha^2$ dynamo and by Blackman \& Brandenburg (2002) for the
$\alpha\Omega$ dynamo. The main conclusion is that for large magnetic Reynolds
number the large scale magnetic field grows first exponentially such that \Eq{helcons}
is obeyed at all times. This behavior could not have been reproduced with an
$\alpha$-effect that is not explicitly time-dependent, such as
for example \Eq{catastrophic}. The exponential growth is then followed
by a resistively slow saturation phase, just like in the simulations of B01.

\begin{figure}[t!]\centering\includegraphics[width=0.5\textwidth]{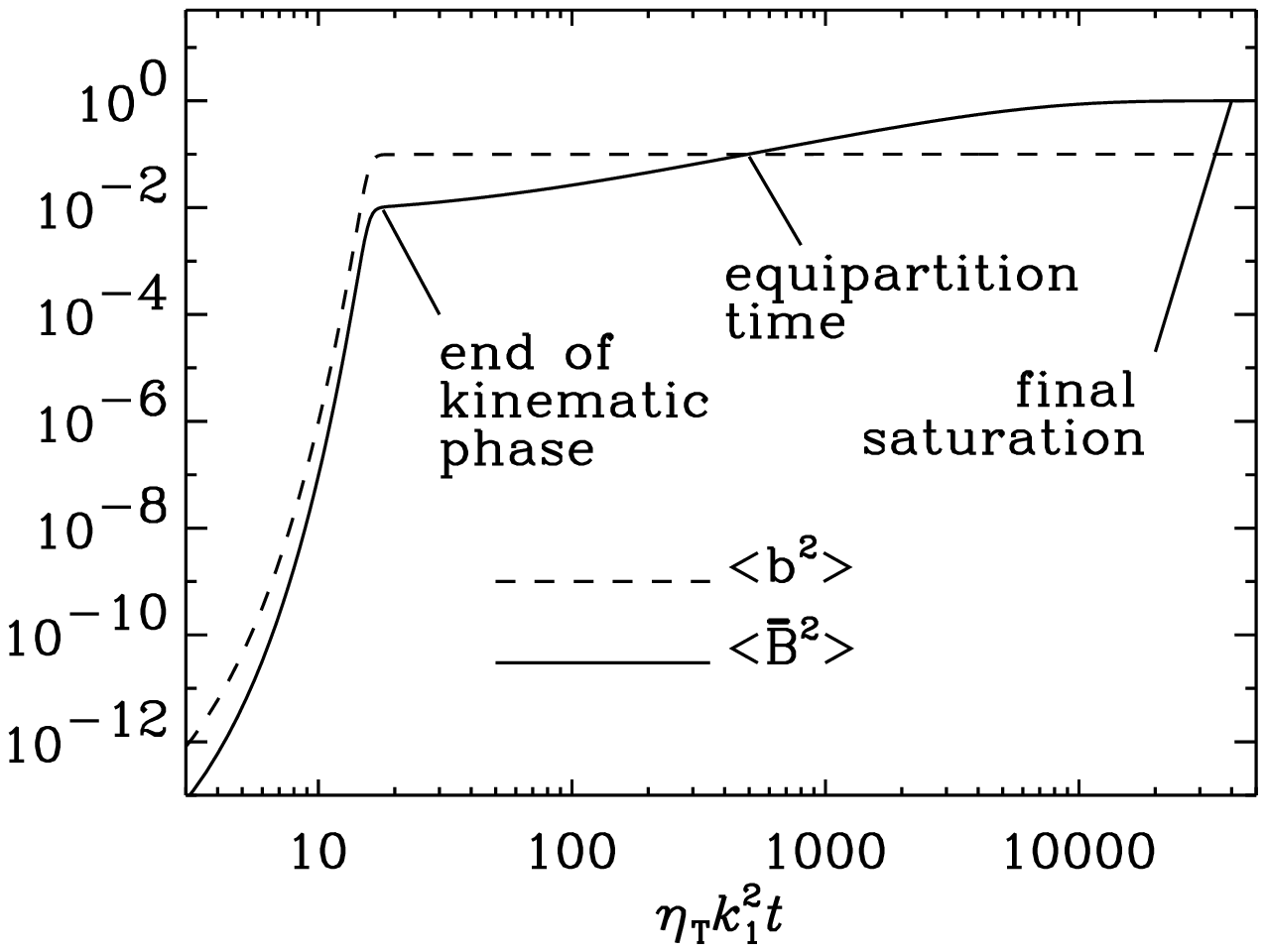}\caption{
Evolution of $\bra{\meanBB^2}$ and $\bra{\bb^2}$ (solid and dashed lines,
respectively) in a doubly-logarithmic plot for an $\alpha^2$ dynamo with
$\eta_{\rm t}=\mbox{const}$ for a case with $k_{\rm f}/k_1=10$.  Note the
abrupt initial saturation after the end of the kinematic exponential
growth phase with $\bra{\meanBB^2}/\bra{\bb^2}\sim0.1$, followed by
a slow saturation phase during which the field increases to its final
superequipartition value with $\bra{\meanBB^2}/\bra{\bb^2}\sim10$.
(Adapted from Blackman \& Brandenburg 2002.)
}\label{Fpcrossing}\end{figure}

The reason there is this slow saturation phase is that \Eq{helcons} is
incompatible with a steady state, where the right hand side of
\Eq{helicity_eqn} must vanish, i.e.\
\EQ
\bra{\JJ\cdot\BB}\equiv
\bra{\meanJJ\cdot\meanBB}+\bra{\jj\cdot\bb}=0
\quad\mbox{(steady state)}.
\label{helcons_steady}
\EN
Since the large scale field is helical, \Eq{JBm} applies and
we have $k_1\bra{\meanBB^2}=k_{\rm f}\bra{\bb^2}$, so the
large scale field {\it exceeds} the small scale field by a
factor $k_{\rm f}/k_1$. In the simulations of B01,
$\bra{\meanBB^2}/\bra{\bb^2}$ was about 4, which is indeed close to
$k_{\rm f}/k_1$=5. In the beginning of the nonlinear regime, \Eq{helcons}
predicts, instead, that $\bra{\meanBB^2}/\bra{\bb^2}$ equals
$k_1/k_{\rm f}$, which was 25 times smaller in B01.
\FFig{Fpcrossing} shows this quantitatively
by solving \Eqs{fullset1}{fullset2}; see
Blackman \& Brandenburg (2002).

In order to bring the field ratio from 1/5 to 5 we have to remove small
scale magnetic helicity resistively. The question is of course what
happens if one considers the effects of boundaries (both at the equator
and at the outer surface): can boundaries remove small scale magnetic
helicity so that the large scale field can saturate at a higher level?
This possibility was first brought up by Blackman \& Field (2000) and
Kleeorin \ea (2000).

\section{From closed to open boxes}

When the restriction to closed or periodic boxes is relaxed, there can be
a flux of magnetic helicity through the surface, so \Eq{helicity_eqn}
has then an additional term,
\EQ
{\dd H\over\dd t}=-2\eta\mu_0C-Q,
\label{helicity_eqn_flux}
\EN
where $H$ and $C$ are magnetic and current helicities, respectively, and
$Q$ is the surface integrated magnetic helicity flux. In the presence of
open boundaries, however, $H$ and $Q$ are no longer invariant under the
gauge transform $\AAA\rightarrow\AAA'+\nab\phi$. We use therefore the
relative magnetic helicity (Berger \& Field 1984, Finn \& Antonsen 1985),
\EQ
H=\int_V(\AAA+\AAA_{\rm P})\cdot(\BB-\BB_{\rm P})\,\dd V,
\EN
where $\BB_{\rm P}$ is a potential field used as reference field within
the volume $V$, and $\AAA_{\rm P}$ is its vector potential. Both $\AAA$
and $\AAA_{\rm P}$ can have different (arbitrary) gauges.

\begin{figure}[t!]\centering\includegraphics[width=0.5\textwidth]{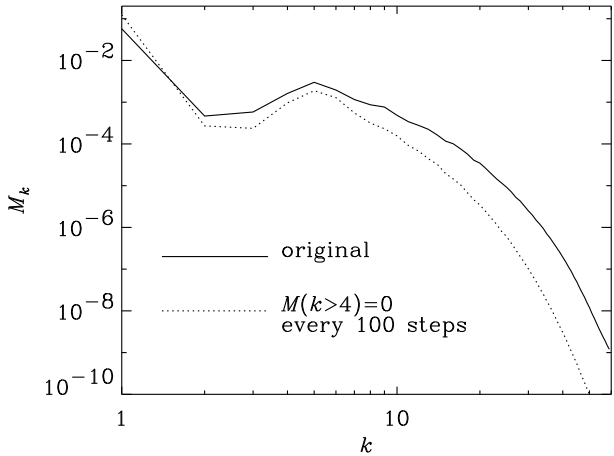}\caption{
The magnetic energy spectrum for the closed box simulation (solid line)
compared with the case where small scale magnetic energy is removed
every 100 time steps, corresponding to
$\delta t=1.3\times(k_{\rm f}u_{\rm rms})^{-1}$. Note that the large
scale magnetic energy (at $k=1$) is enhanced relative to the reference run, whilst
the small scale energy is (as expected) reduced.
}\label{Fpspec_comp}\end{figure}

Following Berger \& Field (1984), the reference field obeys the boundary
condition $\BB_{\rm P}\cdot\nnn=\BB\cdot\nnn$, i.e.\ the normal components
of both fields agree on the boundary. In slab geometry, however, the
horizontally averaged mean field has to be treated separately and the
corresponding reference field is (Brandenburg \& Dobler 2001)
\EQ
\meanBB_{\rm P}=\bra{\meanBB}=\mbox{const},
\EN
so $\meanAA_{\rm P}$ is just a linear function of $z$,
$\meanAA_{\rm P}=-\zzz\times\meanBB_{\rm P}$.
Brandenburg \& Dobler (2001) found from their simulations that most of
the magnetic helicity is lost on large scales, where the sign agrees with
that of the large scale magnetic helicity. This was a bit disappointing,
because one would have hoped that the loss term in \Eq{helicity_eqn_flux}
might supersede resistive losses at small scales. That small scale
losses can at least in principle enhance the large scale field was shown
in subsequent simulations (Brandenburg, Dobler \& Subramanian 2002,
hereafter BDS; see also \Fig{Fpspec_comp}).
The hope is now that this behavior can eventually be demonstrated using
more realistic geometries.

\section{From boxes to spheres}

Spherical geometry is necessary to assess more realistically the helicity
transfer through equator and outer surfaces, and the relative contributions
from rotation, shear, $\alpha$-effect, and turbulent magnetic diffusion.
In a recent paper, Berger \& Ruzmaikin (2000) estimated that the overall
helicity flux at the surface would be around $10^{47}\Mx^2$ per 11 year
cycle. This value was also confirmed by BDS, who computed numerically
solutions of the mean-field dynamo equations in spherical geometry.
The relative magnetic helicity for an axisymmetric mean field,
$\meanBB=b\vec{\hat\phi}+\nab\times(a\vec{\hat\phi})$, takes the very
simple form (BDS)
\EQ
H_{\rm N}=2\int_{\rm N}ab\,\dd V
\label{BF84integral}
\EN
where N denotes the volume of the northern hemisphere. The corresponding integrated
helicity transfer through the outer surface or the equator is
\EQ
Q_{\rm N}=-2\oint_{\partial V}[(\meanemf+\meanUU\times\meanBB)
\times(a\vec{\hat\phi})]\cdot\dd\SSS,
\EN
where $\meanUU=r\sin\theta\,\Omega\vec{\hat\phi}$ is the mean
toroidal flow. Note that $Q_{\rm N}=Q_{\rm NS}+Q_{\rm Eq}$,
where $Q_{\rm NS}$ is the contribution from the outer surface
integral and $Q_{\rm Eq}$ is that from the equatorial plane. 

It is remarkable that even in the presence of just uniform rotation alone
there is a magnetic helicity flux. For a decaying dipolar magnetic field, the
magnetic helicity flux through the equator (from north to south),
or, what is the same, into the outer surface of the northern hemispheres,
or out of the surface at the southern hemisphere, is $\Phi^2$ per rotation,
where $\Phi$ is the magnetic flux through one hemisphere.

\begin{figure}[t!]\centering\includegraphics[width=0.5\textwidth]{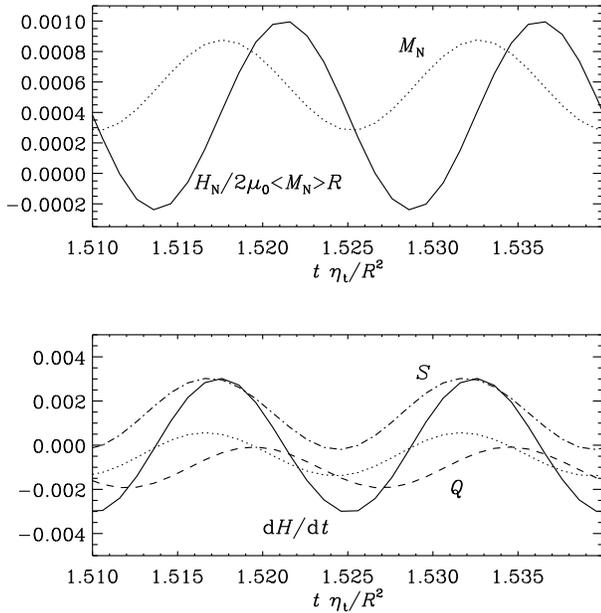}\caption{
Time series of the nondimensional ratio $H/2\mu_0\bra{M}R$ compared with
magnetic energy (in nondimensional units and scaled by 1/20).
Magnetic helicity is mostly positive in the northern hemisphere.
The helicity production, $S$, from $\alpha$-effect and turbulent
magnetic diffusion is mostly positive and balanced here by a
mostly positive magnetic helicity flux (dashed and dotted lines
refer to the contributions from the angular velocity and the
$\alpha$-effect, respectively).
}\label{Fn_an}\end{figure}

However, once the field is sustained by a dynamo effect at a constant amplitude, the helicity
flux must be balanced by the electromotive force (averaged over one cycle), so
\EQA
\begin{array}{lll}
Q&=&2\bra{\meanemf\cdot\BB}V-2\eta\mu_0\bra{\meanJJ\cdot\meanBB}V\\
&\equiv&2\alpha\bra{\meanBB^2}V-2\eta_{\rm T}\mu_0\bra{\meanJJ\cdot\meanBB}V,
\end{array}
\ENA
where $\eta_{\rm T}=\eta+\eta_{\rm t}$ is the total (microscopic
plus turbulent) magnetic diffusivity. Assuming that the dynamo is
saturated by a reduction of the residual helicity, see \Eq{residual_alpha}, $\alpha_0$ and
$Q_{\rm N}$ must have opposite signs. This is because saturation requires that
$\sgn(\alpha_0)=-\sgn(\bra{\jj\cdot\bb})$, but steady state of \Eq{dabdt}
requires that $-\sgn(\bra{\jj\cdot\bb})=\sgn(\bra{\meanemf\cdot\meanBB})$, and
$\sgn(\bra{\meanemf\cdot\meanBB})=\sgn(Q)$. This can also be seen from a
time series of magnetic helicity, and the different terms on the
right hand side the $\dd H_{\rm N}/\dd t$ equation; see \Fig{Fn_an}.

Thus, if $Q_{\rm N}$ (which denotes
only the contributions from the large scale field), is to be identified with
the observed negative magnetic helicity flux found on the solar surface
(Berger \& Ruzmaikin 2000, DeVore 2000, Chae 2000), then we must conclude
that $\alpha$ is negative on the northern hemisphere. This scenario,
where the main magnetic helicity flux results from the large scales,
is consistent with the simulations of BD01. On the other hand, if small
scale magnetic helicity is lost preferentially at small scale, and
the sign of the small scale magnetic helicity is opposite, then $\alpha$
would in that scenario be positive on the northern hemisphere.
This would be consistent with the observed negative sign of current
helicity on the northern hemisphere (Seehafer 1990, Pevtsov \ea 1995,
Bao \ea 1999, Pevtsov \& Latushko 2000), which is plausibly a proxy
of small scale magnetic helicity.

Within the framework of mean-field $\alpha\Omega$ dynamo theory,
a negative $\alpha$ in the nothern hemisphere
would explain the observed migration
of the sunspot belts, so one would not need to resort to
meridional circulation driving the dynamo wave. However,
there is as yet no well established mechanism to explain
a negative $\alpha$ (except perhaps magnetic buoyancy with
shear; cf.\ Brandenburg 1998).

Observations do not seem to be able to tell us which of the two
scenarios is right, because it is difficult to tell whether the
observed magnetic helicity flux is from large or small scale fields.
If the observed magnetic helicity flux is from small scale, one
might wonder why one cannot see the magnetic helicity flux from
the large scales. On the other hand, if the observed magnetic
helicity flux is actually already due to the large scales, one
might expect to see small scale magnetic helicity fluxes at
higher resolution in the future.

\section{Conclusions}

Magnetic helicity seems to play a much more prominent role than
what has been anticipated until recently. It has become clear that
$\alpha$ must satisfy an explicitly time-dependent equation.
The dynamical $\alpha$-quenching theory has significant predictive
power: it describes the different quenching behaviors for helical
and nonhelical fields, the value of the magnetic Reynolds number
is explicitly incorporated, and the magnetic helicity equation
is satisfied exactly at all times. So far, no departures between
this theory and the simulations have been found. A major restriction
of the theory in its present form is however the inability to handle cases
with spatially nonuniform $\alpha$-effect.

\acknowledgements
Use of the PPARC supported supercomputers in St Andrews and Leicester (UKAFF)
is acknowledged.


\vfill\bigskip\noindent{\it
$ $Id: paper.tex,v 1.16 2002/06/11 15:35:54 dobler Exp $ $}

\end{document}